\documentclass[twocolumn]{aastex631}

\received{\today}

\submitjournal{\apj}

\shorttitle{Effects of Early-Forming Massive Stars}
\shortauthors{Lewis et al.}
\graphicspath{{./}{figures/}}

\begin{document}

\title{Early-Forming Massive Stars Suppress Star Formation and Hierarchical Cluster Assembly}

\correspondingauthor{Sean C. Lewis}
\email{sean.phys@gmail.com}

\author[0000-0003-4866-9136]{Sean C. Lewis}
\affiliation{Department of Physics, Drexel University, Philadelphia, PA, USA}

\author[0000-0001-9104-9675]{Stephen L. W. McMillan}
\affiliation{Department of Physics, Drexel University, Philadelphia, PA, USA}

\author[0000-0003-0064-4060]{Mordecai-Mark Mac Low}
\affiliation{Department of Astrophysics, American Museum of Natural History, New York, NY, USA}
\affiliation{Department of Astronomy, Columbia University, New York, NY, USA}
\affiliation{Department of Physics, Drexel University, Philadelphia, PA, USA}

\author[0000-0002-6116-1014]{Claude Cournoyer-Cloutier}
\affiliation{Department of Physics \& Astronomy, McMaster University, Hamilton, ON, Canada}

\author[0000-0001-5972-137X]{Brooke Polak}
\affiliation{Universit\"{a}t Heidelberg, Zentrum f\"{u}r Astronomie, Institut f\"{u}r Theoretische Astrophysik, \\Albert-Ueberle-Stra{\ss}e 2, 69120 Heidelberg, Germany}
\affiliation{Department of Astrophysics, American Museum of Natural History, New York, NY, USA}

\author[0000-0002-3001-9461]{Martijn J. C. Wilhelm}
\affiliation{Leiden Observatory, Leiden University, P.O. Box 9513, 2300 RA Leiden,
The Netherlands}

\author[0000-0003-3483-4890]{Aaron Tran}
\affiliation{Department of Astronomy, Columbia University, New York, NY, USA}

\author[0000-0003-3551-5090]{Alison Sills}
\affiliation{Department of Physics \& Astronomy, McMaster University, Hamilton, ON, Canada}

\author[0000-0001-5839-0302]{Simon Portegies Zwart}
\affiliation{Leiden Observatory, Leiden University, P.O. Box 9513, 2300 RA Leiden,
The Netherlands}

\author[0000-0002-0560-3172]{Ralf S. Klessen}
\affiliation{Universit\"{a}t Heidelberg, Zentrum f\"{u}r Astronomie, Institut f\"{u}r Theoretische Astrophysik, \\Albert-Ueberle-Stra{\ss}e 2, 69120 Heidelberg, Germany}
\affiliation{Universit\"{a}t Heidelberg, Interdisziplin\"{a}res Zentrum f\"{u}r Wissenschaftliches Rechnen, \\Im Neuenheimer Feld 205, 69120 Heidelberg, Germany}

\author[0000-0003-2128-1932]{Joshua E. Wall}
\affiliation{Department of Physics, Drexel University, Philadelphia, PA, USA}

\begin{abstract}
Feedback from massive stars plays an important role in the formation of star clusters. Whether a very massive star is born early or late in the cluster formation timeline has profound implications for the star cluster formation and assembly processes. We carry out a controlled experiment to characterize the effects of early-forming massive stars on star cluster formation. We use the star formation software suite \texttt{Torch}, combining self-gravitating magnetohydrodynamics, ray-tracing radiative transfer, $N$-body dynamics, and stellar feedback to model four initially identical $10^4$ M$_\odot$ giant molecular clouds with a Gaussian density profile peaking at $521.5 \mbox{ cm}^{-3}$. Using the \texttt{Torch} software suite through the \texttt{AMUSE} framework we modify three of the models to ensure that the first star that forms is very massive (50, 70, 100 M$_\odot$).
Early-forming massive stars disrupt the natal gas structure, resulting in fast evacuation of the gas from the star forming region. The star formation rate is suppressed, reducing the total mass of stars formed.  Our fiducial control model without an early massive star has a larger star formation rate and total efficiency by up to a factor of three and a higher average star formation efficiency per free-fall time by up to a factor of seven.
Early-forming massive stars promote the buildup of spatially separate and gravitationally unbound subclusters, while the control model forms a single massive cluster. 

\end{abstract}



\section{Introduction} \label{sec:intro}
The process of star cluster formation involves a wide range of physical processes that remain not entirely understood. Reviews of the field include \citet{mac_low_control_2004}, \citet{mckee_theory_2007}, \citet{portegies_zwart_young_2010}, \citet{klessen_physical_2016}, \citet{krumholz_star_2019}, \citet{girichidis_physical_2020}, and \citet{krause_physics_2020}. 

A star cluster requires millions of years to form and is deeply embedded in dense gas and dust for a significant portion of that time \citep{lada_embedded_2003,chevance_molecular_2020}. Therefore, relying on observations to understand the formation process remains difficult. Computational models provide essential insights to this process. These models have established that cloud properties and galactic influences strongly regulate the conversion of gas into stars within clouds undergoing hierarchical collapse.
These include the turbulent velocity field \citep{ostriker_kinetic_1999,klessen_gravitational_2000}, magnetic field strength and orientation \citep{mckee_dynamical_1999}, gas density profile \citep{chen_effects_2021}, the multiphase nature of the interstellar medium \citep[ISM;][]{ostriker_regulation_2010}, galactic mergers \citep{dobbs_formation_2020}, the galactic gravity field \citep{li_star_2019}, and galactic jets \citep{mandal_impact_2021}.

The feedback from massive stars probably dominates the self-regulation of the star formation process. Computational models have shown that massive stellar feedback including ionizing radiation \citep{matzner_role_2002, dale_ionizing_2012}, non-ionizing radiation \citep{howard_universal_2018}, stellar winds \citep{dale_before_2014, rahner_winds_2017}, and supernovae \citep{rogers_feedback_2013, smith_supernova_2018} can disrupt the parental giant molecular clouds (GMCs) and shut down star formation. For a general review of feedback models employed in many current star cluster formation simulations, see \citet{dale_modelling_2015}. Without these mechanisms, the gravitational collapse of the cloud would continue unimpeded, converting all of the natal gas into stars, in stark contrast to observations of such regions \citep{ostriker_regulation_2010,chevance_life_2022}. 
 
Massive stellar feedback is also thought to regulate sub-cluster structure and assembly. The hierarchical assembly of clusters has been both observed \citep{bressert_spatial_2010, longmore_formation_2014, gouliermis_hierarchical_2017} and demonstrated computationally \citep{maschberger_properties_2010, howard_universal_2018, grudic_top_2018, vazquez-semadeni_hierarchical_2017, vazquez-semadeni_global_2019, chen_effects_2021, dobbs_formation_2022, guszejnov_cluster_2022}. Gas evacuation (via stellar feedback) is crucial to the completion of the assembly process \citep{grudic_top_2018,krause_physics_2020}. In addition, it has been established that \emph{how} the gas is removed from a cluster can potentially affect the cluster structure \citep{smith_infant_2013}. Rapidly evacuated gas can result in cluster destruction or dissolution through the unbinding of stars \citep{lada_embedded_2003,portegies_zwart_young_2010, banerjee_how_2017}. \citet{gavagnin_star_2017} also found weak correlation between feedback strength and unbinding of stars and \citet{li_disruption_2019} saw some dispersal of stars at the highest feedback levels in their parameter study. However, there has been little research as to the effects of \emph{when} gas removal occurs. With our computational model, we test the effects of early-forming massive stars on cluster formation and the hierarchical cluster assembly process.
 
Massive star feedback mechanisms have been shown to slow star formation and contribute to the destruction of the natal cloud. In order to accurately model star cluster formation, each feedback mechanism must be modeled simultaneously within the same computational model. Doing so at the appropriate level of sophistication provides a more realistic star cluster formation framework from which simulations can be constructed. Several recent efforts have created such a framework by combining multiple massive star feedback mechanisms with magnetohydrodynamical (MHD) solvers  \citep{rogers_feedback_2013, dale_before_2014,lancaster_star_2021,grudic_starforge_2021}. Using the \texttt{AMUSE} framework \citep{portegies_zwart_multiphysics_2009, portegies_zwart_multi-physics_2013, pelupessy_astrophysical_2013, portegies_zwart_astrophysical_2018}, we have constructed a hybrid N-body and MHD simulation environment for modeling cluster formation called \texttt{Torch} \citep{wall_collisional_2019,wall_modeling_2020}. 
We combine stellar evolution, massive stellar radiative feedback, winds, and supernovae into an adaptive mesh MHD framework and couple this with high precision $N$-body dynamics, allowing us to follow the dynamics of individual stars within an actively forming cluster that exposes the gas to feedback from the massive stars. We test the hypothesis that the timing of massive star formation plays a vital role in the star formation and star cluster assembly processes because once formed, massive stars disrupt the natal gas cloud, limit global star formation efficiency (SFE), and promote the formation of stellar subclusters while hindering their assembly into a young massive cluster.

We test the impact of early-forming massive stars by comparing simulations with identical initial conditions but varying masses for the first formed star, randomly choosing it from the initial mass function (IMF) in our fiducial run, or forcing it to have a mass of 50, 70, or 100~M$_\odot$.

We describe our simulation initial conditions and parameter space in Section \ref{sec:methods}. We analyze the effects of early-forming massive stars on the gas and star cluster formation in Section \ref{sec:analysis}. We discuss our results, compare them to previous works and note the limitations of our model in Section \ref{sec:discussion}, and finally conclude in Section \ref{sec:conclusion}. 

\section{Methods} \label{sec:methods}

\texttt{Torch}\footnote{\url{https://bitbucket.org/torch-sf/torch/src/main} using commit \texttt{ 811d35ea069ca4a7e099e62bb4f0580f0a49cf29} for runs presented in this paper.} \citep{wall_collisional_2019} couples the adaptive mesh MHD code \texttt{FLASH} \citep{fryxell_flash_2000}, including modules implementing heating and cooling, ray-traced radiative transfer \citep{baczynski_fervent_2015}, and sink particle creation module \citep{federrath_modeling_2010-1} with the \texttt{AMUSE} framework. 
Within \texttt{AMUSE}, we also use the $N$-body dynamics solver \texttt{ph4} \citep{mcmillan_simulations_2012}, the binary and close encounter modules \texttt{multiples} \citep{portegies_zwart_astrophysical_2018} and \texttt{smalln} \citep{hut_building_1995,mcmillan_binary--single-star_1996}, as well as the stellar evolution module \texttt{SeBa} \citep{portegies_zwart_population_1996}.

\texttt{FLASH} is integrated into \texttt{AMUSE} using the hierarchical coupling strategy \citep{portegies_zwart_non-intrusive_2020}, which is a generalization of the gravity bridge scheme developed by \citet{fujii_bridge_2007}.
With this coupling, we are able to model self-gravitating, radiatively heated and cooled, magnetized GMCs, while also forming stars from the gas and resolving individual stellar dynamics. Within FLASH, we use an HLLD Reimann solver \citep{miyoshi_multi-state_2005} with order 3 PPM reconstruction \citep{colella_piecewise_1984}. If a particularly strong shock occurs that triggers numerical instability, we briefly switch to the more diffusive HLL solver \citep{einfeldt_godunov-type_1991} with first order Godunov reconstruction \citep{godunov_finite_1959}.
Our method of converting collapsing gas into stars uses \texttt{FLASH}'s sink particle module \citep{federrath_modeling_2010-1} which replaces Jeans unstable gas \citep{truelove_jeans_1997-1} with a sink particle (sink from here on) with a mass equivalent to the replaced gas (see \citealt{federrath_modeling_2010-1} for details on sink creation and accretion criteria). 

The adaptive mesh is required to refine such that the \citet{truelove_jeans_1997-1} length is resolved by 4 or more cells. 
The computational domain is a cube of size 17.5 pc. At the top refinement level, the entire grid is represented by a single block of $16^3$ cells. Each successive refinement level can break a block up into four smaller blocks. Our runs have a maximum refinement level of 3, on which cells are 0.27 pc on a side. The outer edges of the computational domain are governed by outflow boundary conditions to allow gas to properly escape from the star forming region.

We evolve four simulations with identical initial conditions. In the first, we treat the simulation as a standard \texttt{Torch} run by with the usual physical prescriptions. In the other runs, the first star to be born to have a mass of 50, 70, or 100~M$_\odot$ (referred to as the 50M, 70M, and 100M runs respectively). 
We evolve the simulations until star formation ceases or the forced massive star goes supernova ($\sim6$~Myr total or 4 Myr after star formation begins). In the case of the 100M run, nearly all of the gas and stars eventually escape the computational domain. This allows for large timesteps and a simulation that extends further in time than the other runs.

\subsection{Initial Conditions}

We initialize a $10^4$~M$_{\odot}$ GMC as a spherical cloud with a radius of 7.25 pc and with a Gaussian density distribution \citep{bate_modelling_1995-1} with a standard deviation of 4.89 pc. The cloud is also surrounded by a low density background medium. All gas has solar metallicity, which remains constant throughout the simulations. No background galactic gravitational potential is applied.
We initialize the cloud and background medium to be in pressure and thermal equilibrium and choose the gas densities and temperature accordingly. The spherical cloud is in the cold neutral medium phase with the central gas density being 521.5 cm$^{-3}$ at a temperature of 20.6 K. The cloud edge density is 1/3 the central density. The surrounding low-density, higher-temperature gas is in the warm neutral medium phase with density 1.3 cm$^{-3}$ and temperature 6105.3 K.
We apply a turbulent Kolmogorov velocity distribution (R. W\"unsch 2015, personal communication) to the dense gas such that it is subvirial with a virial ratio $\alpha = 2T/|U| = 0.12$, where $T$ is the total kinetic energy and $|U|$ is the magnitude of the potential energy. Low virial parameter clouds are appropriate for regions containing massive star formation \citep{kauffmann_low_2013} and such low virial parameter clouds have been catalogued \citep{roman-duval_physical_2010, wienen_ammonia_2012}.
A subvirial initial state provides an environment where star formation occurs readily in dense regions and tends to reduce the effect of early-forming massive stars due to the reduced penetrating power of the feedback mechanisms through the dense gas.
We initialize the cloud with a uniform $3\,\mu G$ magnetic field along the z-axis and allow the gas turbulence to mix the field, mirroring the setup defined by \citet{wall_collisional_2019,wall_modeling_2020}.\footnote{Runs M3f and M3f2 in \citet{wall_modeling_2020} did not include magnetic fields due to an incorrect initialization procedure; this oversight has been corrected so all our runs begin with the uniform magnetic field.} 
We also apply a background far ultraviolet radiation field with a constant flux of 1.7$G_{0}$ \citep{draine_photoelectric_1978}, where $G_{0}=1.6\times10^{-3}$ ergs s$^{-1}$ is the \citet{habing_interstellar_1968} flux. We estimate the extinction of the far ultraviolet flux using the local Jeans length \citep{truelove_jeans_1997-1}. We also assume a constant gas ionization rate due to cosmic rays $\zeta=10^{-17}$s$^{-1}$.

\subsection{Sink Properties and Star Formation}

As the dense gas collapses, sinks may form at the highest levels of refinement if the conditions detailed in Section 2.2 of \citet{federrath_modeling_2010-1} are satisfied. Sinks in our simulation have accretion radii of 0.67 pc. They are fixed to grid cell centers and move at each time step to the cell of lowest gravitational potential within their accretion radii. Sinks also merge together if their radii overlap.

Once a sink forms, we assign to it a list of random stellar masses sampled from the \citet{kroupa_initial_2002} IMF with a maximum of 150 M$_\odot$ and a minimum of 0.08 M$_\odot$, following \citet{weidner_maximum_2006}, \citet{sormani_simple_2017-1} and \citet{wall_collisional_2019}. Once a sink has accreted enough material to match or exceed the mass of the next star to be formed, a non-accreting star particle (star from here on) is placed on the grid inside of the sink radius and the same mass is removed from the sink. The initial position of a new star is randomly sampled from a spherical Gaussian distribution positioned at the center of the sink. The initial velocity components of a star are sampled from a Gaussian distribution with a scale set to the speed of sound of the gas on which its parent sink is sitting.
The sink can then continue to accrete gas and the star is permitted to move throughout the computational domain under the gravitational influences of the gas, sinks, and other stars. Sink accretion may continue until the local gas reservoir is exhausted, or the gas is heated by ionization or shocks from a massive star.

Stars are not tied to the structure of the computational mesh, but do exert gravitational forces on and experience gravitational forces from the gas as well as other stars and sinks. Stars start at zero-age main sequence (ZAMS). Stars above 7 M$_{\odot}$ produce feedback effects in the forms of photoelectric heating and ionizing fluxes, stellar winds, and supernovae. These feedback mechanisms are injected into the computational domain based on the star's evolution as modeled in \texttt{SeBa}.\footnote{We updated the time step determination process in \texttt{Torch} to ensure massive stars take small enough \texttt{SeBa} evolution steps compared to the current gas dynamical timestep to resolve evolutionary changes in their properties.}
Since stars are placed at ZAMS, massive stars begin their feedback the instant they are placed onto the grid. We therefore do not resolve any earlier feedback effects that take place during the massive star accretion phase. \texttt{SeBa} also models the deaths of massive stars, taking stellar mass, end-of-life composition, and metallicity into account when determining supernova types. Only the early-forming massive stars reach the supernova stage in our runs. They all detonate as supernovae, as their metallicity places them outside the regime of massive stars that directly collapse to a black hole \citep{heger_how_2003}.

\subsection{Early-Forming Massive Stars}
Spawning a massive star in a typical \texttt{Torch} run is a rare event. Each sink has tens of thousands of stars in its stellar mass list but only a few dozen are very massive ($\ge50$~M$_{\odot}$) and they are randomly placed within the list. Therefore, within our computational model it is unlikely yet possible for a sink to have a very massive star as one of its first mass list entries. To explore the effects of the early formation of a very massive star, we force the first-formed sink to have either a 50, 70, or 100 M$_{\odot}$ star as the first entry of its mass list. 
Before the sink begins accreting gas for the 50, 70 or 100 M$_{\odot}$ star, our implementation allows for the formation of 6 stars (the most massive of which is 0.8 M$_\odot$ and an average mass of 0.45 M$_\odot$). 

The chosen parent sink for the very early-forming massive star is also the first sink to form. This was a deliberate choice, as this sink forms near the center of the collapsing cloud, thus has a substantial supply of infalling material to accrete, and so can spawn the very massive star as the first born from the sink. Other sink particles are still able to form elsewhere in the collapsing cloud if the formation conditions are satisfied. Indeed, another sink does form around 6-7 parsecs from and 500 kyr after the parent. In each forced run, this sink is able to form $\sim24$~M$_\odot$ of stars (29 total stars with the most massive being 9.8M$_\odot$) before the early-forming massive star forms.

\section{Analysis} \label{sec:analysis}
\begin{figure*}[!htb]
    \centering
	\includegraphics[width=0.8\linewidth]{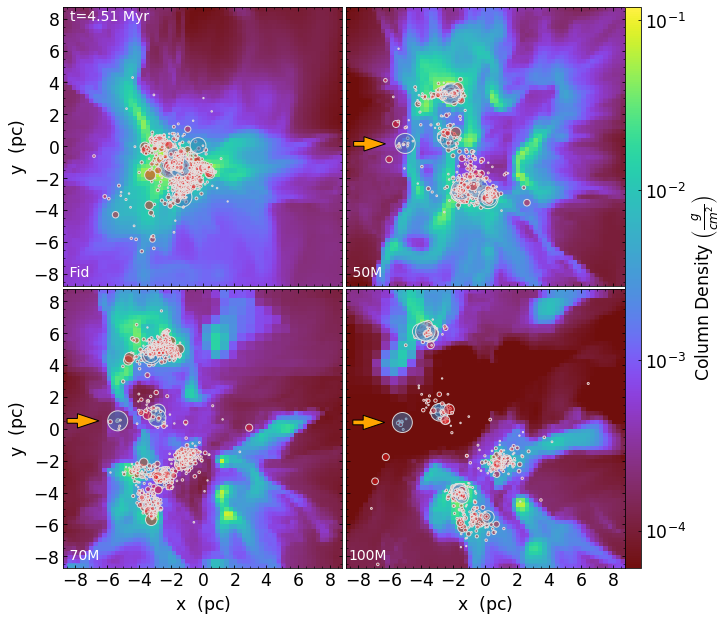}\\
    \caption{Snapshot of the fiducial run and early-forming massive star runs 50M, 70M, and 100M at $2\tau_{\rm ff}$. All runs start from identical initial conditions. The gas column density as well as stars are plotted. Stars denoted by red circles have masses $<7$~M$_\odot$ and do not produce any form of feedback in our model. Stars denoted by blue circles have masses $>7$~M$_\odot$ and produce ionizing radiation, winds, and supernovae. Plotted star sizes are scaled by the star's mass. The orange arrows highlight the forced massive star in each of the forced runs. }
    \label{fig:sim_grid_plot}
\end{figure*}

Visual inspection of the simulations at a characteristic time reveals the increased destructive effects of early-forming massive stars to both the gas and the hierarchical assembly of the resulting star clusters. Figure~\ref{fig:sim_grid_plot} shows the column densities of each run at simulation time $t=4.51 \mbox{ Myr }= 2\tau_{\rm ff}$, where $\tau_{\rm ff}$ is the free-fall time of the initial gas cloud. At this characteristic time, the gas is more fragmented in the runs containing early-forming massive stars (50M, 70M, 100M) with the 100M run showing the most fragmentation. We further choose $2\tau_{\rm ff}$ as the snapshot to analyze since after this time subclusters begin exiting the computational domain.

In the fiducial run, the gas remains present in the central regions of the cloud even at the end of the run. 
The stars formed in the fiducial run collect into a single large cluster while the forced runs result in several localized associations of stars. 

\begin{figure*}[!htb]
	\centering
	\includegraphics[width=0.8\linewidth]{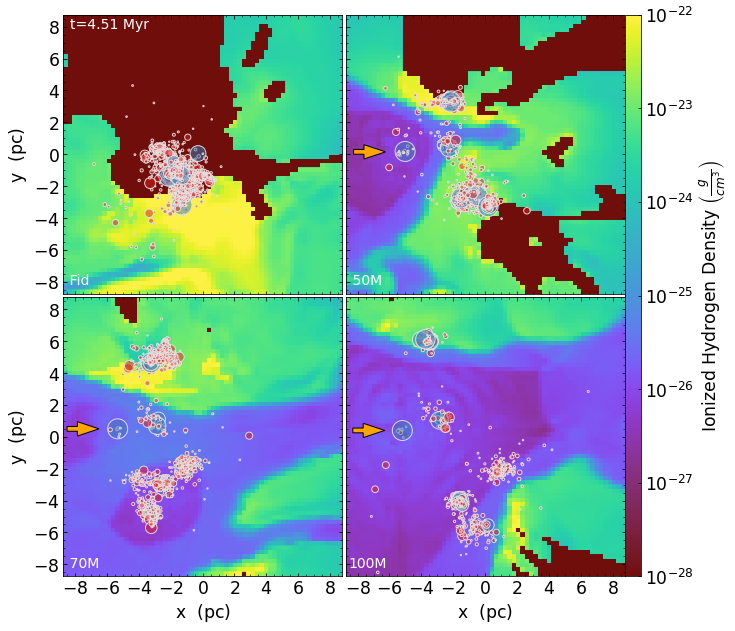}
    \caption{The same snapshot as in Figure \ref{fig:sim_grid_plot} at $2\tau_{\rm ff}$ but as a slice at z=0 pc of ionized Hydrogen density. To highlight the morphology of the ionized regions, a gas cell is determined to be fully ionized if its ionization fraction exceeds 0.99.  Gas cells that do not meet this threshold do not have their densities represented and appear as the dark maroon color. The stars are shown but note that they may lie above or below the slice plane. The orange arrows highlight the forced massive star in each of the forced runs.}
    \label{fig:sim_grid_plot_Hatom}
\end{figure*}

In the forced runs, since the sink selected to form the early massive star was close to the center of mass of the cloud, the early-forming massive star also forms near the center of mass. Though the parent sink does pick up a drift velocity of a few kilometers per second as it falls into the cloud, once the early massive star forms, it rapidly evacuates the cloud center of gas, thereby removing the gas potential as well. As a result, the massive star and its parent sink (in all three of the forced runs) continues to drift as the simulation progresses. The early-forming massive stars at $2\tau_{\rm ff}$ (about 2.5 Myr after their formation) can be seen in Figures \ref{fig:sim_grid_plot} and \ref{fig:sim_grid_plot_Hatom} indicated by orange arrows. In the time since their formation, they all were able to drift several parsecs from their initial positions (within 1 pc of the simulation origin). The stars that are born after the early-forming massive star still collect in clumped groups but more readily form in gas clumps that build up on the outskirts of the original spherical cloud rather than at the density center as in the fiducial run.

Figure \ref{fig:sim_grid_plot_Hatom} 
shows the density of fully ionized gas as a slice through the computational domain at $z=0$~pc. 
At $2\tau_{\rm ff}$, the early-forming massive stars have ionized a significant portion of the computational domain and their low-density ionized wind bubble is dominant particularly in runs 70M and 100M. Each forced run still has regions of partially or fully shielded gas which have not been fully ionized with the 50M run having the most extensive regions of gas that is not fully ionized.
In the fiducial run a distinct wind bubble is not yet visible as massive stars in the central cluster have not had time to carve away the surrounding dense gas. However, their ionizing radiation has been able to penetrate and fully ionize the dense gas below the cluster, while large portions of the +y quadrants are still partially ionized or shielded neutral gas.

So while the fiducial run has not yet begun to mechanically disrupt the gas surrounding the main cluster, ionizing feedback has still penetrated into portions of the grid. Meanwhile, the 50M, 70M, and 100M runs show extensive ionization throughout the region and clear mechanical disruption via their developing wind bubble with the level of disruption seeming to increase monotonically with increasing forced star mass.

To quantify the effects of early-forming massive stars, we analyze the energetics and behavior of the gas as well as the conversion of the gas into stars. We also quantify early-forming massive stars' effects on the hierarchical assembly of subclusters by identifying clusters and investigating their properties.

\subsection{Gas Disruption and Expulsion} \label{subsec:expulsion}

We measure the total energy of all gas in the computational domain by summing its mechanical energy, thermal energy, magnetic energy, and the gravitational energy of gas on gas and stars on gas. When the total energy becomes positive, the gas is in a globally unbound state. While this does not necessarily indicate that the natal GMC has been destroyed or will entirely dissipate (there still can be regions of gravitationally dominated gas that will continue to collapse and form stars), it does provide a metric to quantify the global effects of energy injection. 

\begin{figure}[!htb]
	\includegraphics[width=\columnwidth]{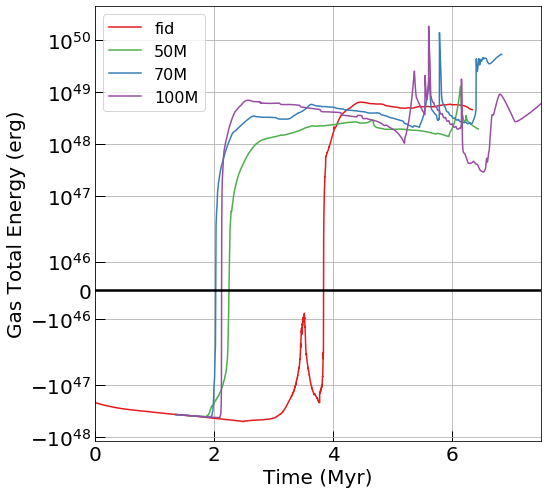}
    \caption{Total energy of gas in the computational domain, comprising kinetic, thermal, and magnetic energy of the gas,  energy from gas self-gravity, and gravitational energy due to stars acting on the gas for the fiducial and forced runs. The $y$-axis is truncated close to zero to reveal detail at times before and after the transition from negative to positive total energy. In each case, the energy is first dominated by gravitational potential energy, and then becomes dominated by gas kinetic energy and gas thermal energy once the massive star forms in the forced runs or several $>$50 M$_{\odot}$ stars form in the fiducial run.}
    \label{fig:gas_energy}
\end{figure}

Figure \ref{fig:gas_energy} shows that the gas reaches a globally unbound state at 2.0--2.1 Myr after simulation start for the forced runs, and at 3.9 Myr for the fiducial run. The transition to a globally unbound state of gas coincides with the formation of massive stars (1.8, 1.9, 2.1 Myr for the 50, 70, 100 M$_{\odot}$ stars respectively) while the fiducial transition occurs during a rapid formation of stars, including two stars with masses over 60 M$_{\odot}$. 

The fiducial run also has a spike in energy a few hundred thousand years before its transition to a positive energy state. This initial spike is driven in part by the formation of a 29 M$_{\odot}$ star several parsecs from the center of mass of the cloud at $t=2.51$~Myr. It remains the most massive star on the grid for the next million years. As the HII region around the star increases in volume, we see an increase in the total thermal energy of the gas in conjunction with a more slowly increasing total kinetic energy produced by feedback from this star. Around $t=3.1$~Myr, several lower mass feedback-producing stars (ranging from 12.49 to 17.24 M$_{\odot}$) form in the center of mass of the collapsing cloud. Their feedback in addition to the influence of the 29 M$_{\odot}$ star results in the brief energy spike at $t=3.5$~Myr.
The total gas energy then declines rapidly as more gas falls towards the gravitational center of the cloud and more stars are formed there. An off-center star such as this cannot markedly reduce gas collapse by sweeping up large amounts of dense cold gas as the wind and ionization bubbles expand, unlike a star closer to the center of mass. 

The gas total energy in the fiducial run finally becomes positive immediately following the formation of a 69.92 M$_{\odot}$ star at 3.76 Myr followed by a 49.86 M$_{\odot}$ star at 3.85 Myr, and a 62.96 M$_{\odot}$ star at 3.96 Myr along with two dozen other less massive feedback injecting stars throughout the same time-span. 

The early-forming massive stars significantly accelerate the global unbinding of the available gas. The large spikes in energy at 5--6~Myr for the forced runs are due to the injection of Wolf-Rayet winds from the early-forming massive star and the subsequent supernova explosion. Since the early-forming massive stars drift significantly during their 4 Myr lifespans, the supernovae occur near the edge of the computational domain in the 50M and 70M runs and the forced star in the 100M run actually leaves the computational domain before going supernova. By the last simulation output the most massive stars in the fiducial run are still 2--3 Myr away from these final stages of stellar evolution. Because of this, the global gas energy of the fiducial run fluctuates far less than in the final stages of the forced runs. 

\begin{figure*}[!htb]
    \centering
	\includegraphics[width=\linewidth]{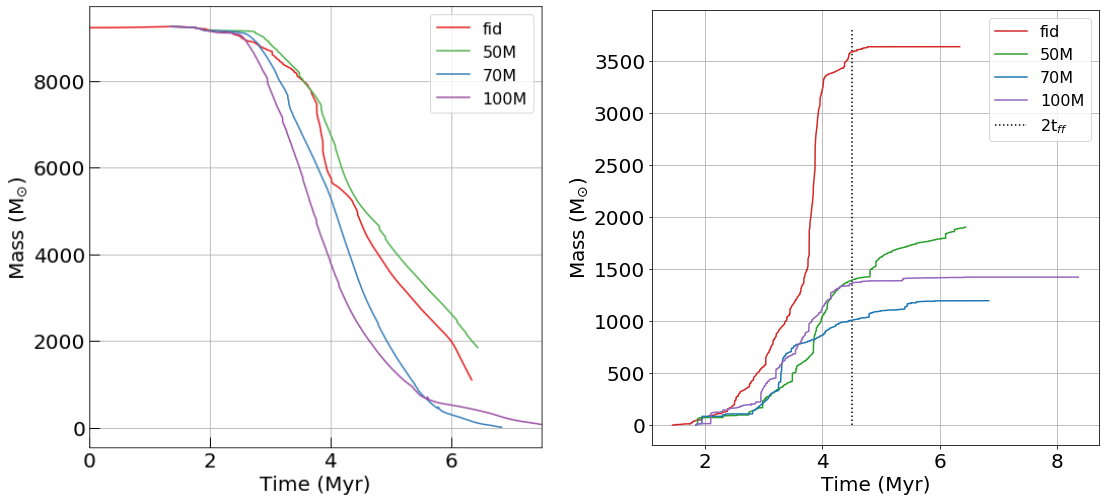}
    \caption{(Left) Total gas mass in computational domain. Reduction in gas is due to gas leaving the domain or accreting onto sinks and ultimately becoming stars. (Right) Cumulative ZAMS stellar mass for the fiducial and three forced massive star runs. The dotted vertical line marks $2\tau_{\rm ff}$, the focus of Sect.~\ref{sec:Hierarchical}.}
    \label{fig:gas_mass_cum_star_mass_combo}
\end{figure*}

The left plot of Figure~\ref{fig:gas_mass_cum_star_mass_combo} shows that the total amount of gas mass in the computational domain decreases most rapidly in the 70M and 100M runs. In each run, gas loss occurs by expulsion from the computational domain or by accretion onto sinks followed by conversion into stars. The right plot of Figure \ref{fig:gas_mass_cum_star_mass_combo}, which shows the cumulative ZAMS stellar mass placed on the grid, demonstrates that the  majority of the decrease in global gas mass in runs 50M, 70M and 100M is due to gas expulsion from the grid. On the other hand, the fiducial run forms significantly more stars than the forced runs and nearly the entirety of the decrease in gas mass before 4.5~Myr is due to the conversion of gas to stars. However after $2\tau_{\rm ff}$, nearly all of the subsequent decrease in global gas mass is due to gas expulsion. 

The 70M and 100M runs both have the majority of stars exit the computational domain by the end of the run. Interestingly, the 70M run cumulatively forms $\sim200$~M$_\odot$ less stellar material than the 100M run. We do not investigate whether there are any unique mechanisms behind this but instead focus on the significant discrepancy between the forced runs and the fiducial run.

\begin{figure}
	\includegraphics[width=\columnwidth]{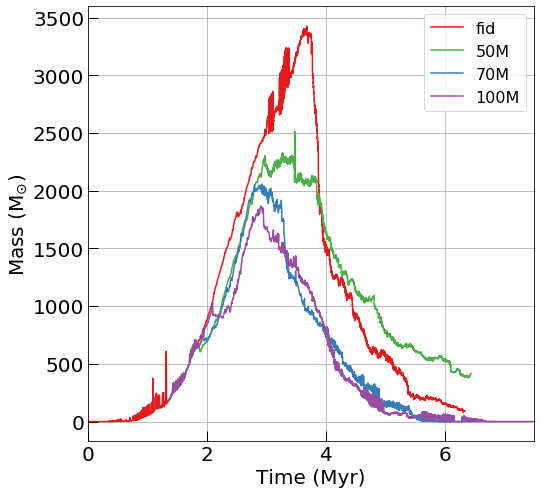}
    \caption{Total mass of gas in the computational domain that satisfies the Jeans criterion, one of the six criteria necessary for sinks to form and one of four sink accretion criteria.}
    \label{fig:jeans_gas}
\end{figure}

The 50M forced run was still actively forming stars at the end of the run even after the 50 M$_{\odot}$ star had gone supernova. Although it has formed many fewer stars than  the fiducial run, 2000 M$_{\odot}$ of gas remains on the grid, so it still has the possibility of forming enough stars to match the fiducial run. However, there are two reasons that it is unlikely that the remaining gas on the grid will produce any stars or subclusters that will interact with the stars on the grid at the time of our analysis. First, as can be seen in Figure \ref{fig:jeans_gas}, only 25\% of the gas remaining on the grid in model 50M meets the Jeans instability criterion, one of the six that must be met to allow gas to be converted into star-forming sinks. 
Second, even if any of the unstable gas can form sinks that then accrete more gas to form stars, Figure~\ref{fig:gas_energy} shows that the gas has positive total energy, and so is unlikely to collapse further. We explore this reasoning more in Section \ref{sec:Hierarchical}.  

The fiducial, 70, and 100 M$_{\odot}$ runs end with long periods without star formation. In the 70 and 100 M$_{\odot}$ runs, nearly all of the initial gas has been exhausted, with only 10--20 M$_{\odot}$ of hot gas remaining.
The fiducial run has nearly all of its stars and all of its sinks remaining in the computational domain. The lack of growth of the total stellar mass in this case is due to the sinks being concentrated at the stellar center of mass of the cloud which becomes entirely devoid of gas after the formation of several high mass stars during the rapid star formation event at 3.8 Myr. There is also a total of 1000~M$_{\odot}$ of gas in the computational domain at the end of the run, none of it is gravitationally collapsing, so star formation has ceased. 

\subsection{Sink Accretion and Star Formation}
\begin{figure}[!htb]
	\includegraphics[width=\columnwidth]{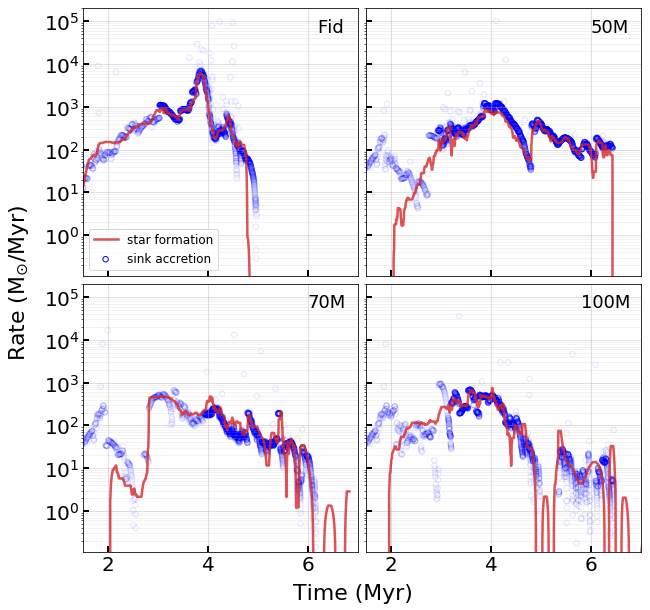}
    \caption{Global sink accretion (\emph{blue circles}) and star formation (\emph{red line}) rates for each run. The SFR line is smoothed with a 100 kyr windowed median to correct for effects of taking the derivative of the spline fit.}
    \label{fig:sink_accretion_rate}
\end{figure}

We show the rate of gas accretion onto all sinks present on the grid as well as the rate of star formation in Figure \ref{fig:sink_accretion_rate}. Accretion onto sinks requires gas to be converging, bound to the sink, Jeans unstable, and within the sink's accretion radius. We expect the total accretion rates to fluctuate as more sinks form, as more gas flows into and onto the sink regions, and as massive stars form that heat or expel dense gas, preventing its accretion. Higher rates of accretion indicate lesser degrees of gas disruption by massive stars. Very low or zero accretion indicates that gas in, on, and around sinks has either been exhausted, or has been disrupted by massive star feedback: heated so that it is no longer Jeans unstable or expelled from the sink accretion zones. The red line in Figure \ref{fig:sink_accretion_rate} tracks the global star formation rate (SFR). We calculate the SFR by taking the derivative of a spline fit applied to the cumulative stellar mass. We then smooth the result using a moving-median filter with a 100 kyr window size to avoid effects due to over-fitting.

Interestingly, the fiducial, 50M, and 100M runs have distinct peaked accretion rates at $\sim$4~Myr (see Fig.~\ref{fig:sink_accretion_rate}), which similarly decrease for the next megayear or so. The fiducial peak is an order of magnitude higher than the median accretion rate of the forced runs, which is clearly represented in the cumulative stellar mass formed as shown in Figure \ref{fig:gas_mass_cum_star_mass_combo}. The violent accretion event in the fiducial run was able to produce enough massive stars to terminate all further accretion in the computational domain. The forced massive stars are able to stifle sink accretion in the time immediately after their formation at $\sim$2~Myr (as seen most prominently in the 50M and 70M runs). At 3 Myr, as the outer regions of the cloud continue to collapse, the accretion rates in the forced runs are able to recover to levels similar to the fiducial run at the equivalent time but the forced runs never experience a similarly rapid gas accretion event. Gas accretion also continues for much longer in the forced runs than the fiducial. The SFR diverges from the sink accretion rate at early times in the forced runs because the early-forming massive star only allows its parent sink to accrete gas until the star forms. Once the massive star forms, and as more sinks form, the SFR closely tracks the sink accretion rate and shows similar trends. 

Some significant outliers in the sink accretion data can be seen. These correspond to accretion of large masses of gas onto a single sink. A sink can move to a nearby region of very dense gas that previously was beyond the sink's accretion radius. If the gas, now within the accretion radius, satisfies all accretion criteria, a large mass of gas accretes in a single timestep, also resulting in high accretion rates. 
This behavior is not unexpected and is easily explained by the movement of sinks onto particularly dense cells of readily accreting gas, although 0.1\% of the GMCs initial mass being accreted in a single timestep likely can be further resolved with a more refined grid structure allowing for more detailed accretion events.

We require the volume within the sink accretion radius to be cold gas with $T <$100~K in order for the sink to form stars. This choice was made as we sample each newly-born star's relative velocity from the sound speed of the gas in the sink volume. If the gas is too hot, stars would be placed with unrealistic relative velocities. Because of this requirement, it is possible for sinks to accrete gas but be unable to form stars if the region within the sink radius is heated by stellar feedback. Such an effect can be seen at the end of the fiducial run when the SFR diverges from the sink accretion rate fiducial run (see Fig. \ref{fig:sink_accretion_rate}). This model can have an undesirable effect of trapping gas within a sink until it ventures into a region of cold gas. Improved models such as preventing sink accretion if star placement is forbidden have been proposed and will be implemented in future versions of \texttt{Torch}.

In each of the simulations there are five sinks present at $2\tau_{\rm ff}$. In the 50M run, there is 110~M$_\odot$ of gas trapped within these sinks, the most massive of which contains the majority of the mass at 96~M$_\odot$. The 70M run has 22~M$_\odot$ trapped with the most massive containing 17~M$_\odot$. The 100M run has just over 3~M$_\odot$ of trapped gas, with 2.5~M$_\odot$ in the most massive sink. The fiducial run has a total of 19~M$_\odot$ trapped in its sinks, with 18~M$_\odot$ in the most massive one.  

The SFE per free-fall time
\begin{equation}\label{eqn:sfe}
\epsilon_{\text{ff}} = \dot{M}_{*}\frac{t_{\text{ff}}}{M_{\text{g}}}
\end{equation}
where $\dot{M}_*$ is the instantaneous star formation rate, $t_{\text{ff}}$ is the cloud initial free-fall time, and  $M_{\text{g}}$ is the remaining gas mass. Averaged over the star formation time,  $\epsilon_{\text{ff}}$ is 0.23, 0.08, 0.03, and 0.04 for the fiducial, 50M, 70M, and 100M runs respectively. These results are consistent with observed values \cite[as collected and reported by][]{krumholz_star_2019}. The high SFE of the fiducial run is likely due to the low virial ratio of the initial cloud which promotes a more aggressive conversion of gas to stars. It is also worth noting that our model assumes a 100\% SFE for gas that a sink accretes, which may result in a higher overall SFE. In addition, our model does not yet include protostellar jets, a disruptive feedback effect that would lower the SFE \citep{federrath_inefficient_2015,appel_effects_2022}. Early massive star formation in our simulations significantly disrupts early gas accretion and prolongs the star formation history of a cloud but at the cost of stifled sink accretion rates as well as greatly reduced star formation rates and average formation efficiencies per free-fall time.
\subsection{Cluster Properties}  \label{sec:Hierarchical} 
\begin{figure*}[!htb]
	\includegraphics[width=0.8\linewidth]{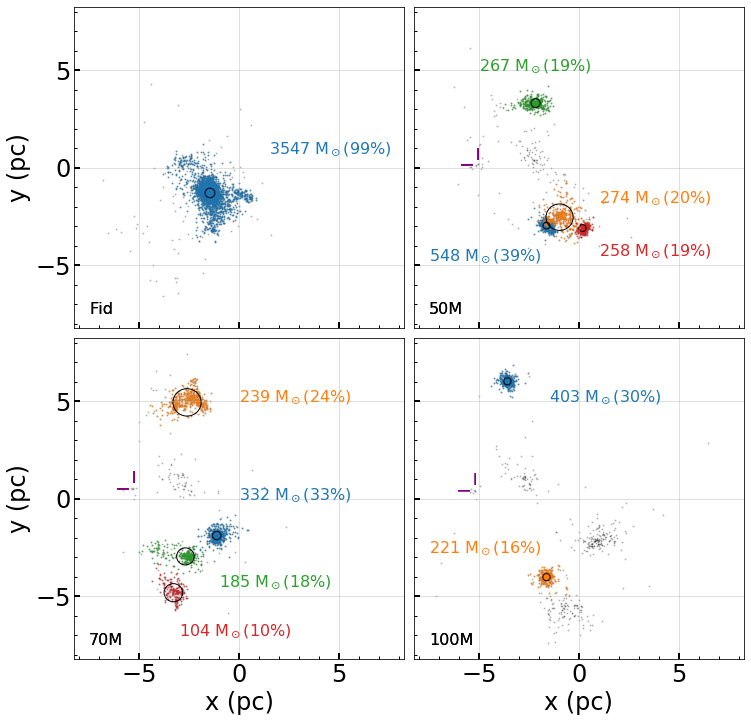}\\
	\centering
    \caption{Snapshot of star clusters formed in fiducial run and forced runs at $2\tau_{\rm ff}$. Clusters are identified using DBSCAN. Each cluster is labeled with color-coded text of cluster total mass and mass fraction of all stars formed contained within the cluster. Overlaid on each cluster is its half-mass radius. In each of the forced runs, the early-forming massive star does not belong to an identified cluster. Stars that are not members of clusters are shown as grey points. The position of the early-forming massive star is indicated with the purple cross-hairs.}
    \label{fig:clust_grid_plot}
\end{figure*}

We identify stellar clusters at $2 \tau_{\rm ff}$ and examine their properties to demonstrate the impact of early massive star formation on the clusters (see Fig.~\ref{fig:clust_grid_plot}). We chose this characteristic time to perform our clustering analysis because some star clusters leave the domain later in the simulations.

\subsubsection{Cluster Identification}

We use \texttt{DBSCAN} from the {\tt scikit-learn} Python library \citep{pedregosa_f_scikit-learn_2011} to identify clusters following \citet{wall_modeling_2020}. \texttt{DBSCAN} identifies clusters as collections of core and boundary particles. We choose core particles to be stars that have at least twelve neighbors within 0.66 pc, while boundary particles lie within 0.66 pc of a core particle but have fewer than twelve neighbors. Our parameters are slightly more restrictive than those used in \citet{wall_modeling_2020}: we set an outer density limit for the maximum distance calculation to be 2.0 M$_\odot$/pc$^3$ rather than 1.0 M$_\odot$/pc$^3$, to better separate clusters from more sparse field stars as well as promoting the separate identification of two nearby clusters whereas a less restrictive model would only identify one.

It is important to note that \texttt{DBSCAN} only groups stars into clusters based on their spatial proximity to one another and does not consider any dynamical or energetic properties of the cluster members.
In order to include some energetic information in the cluster identification process, we check whether each cluster member is bound to the center of mass of its parent cluster. If the fraction of unbound stars in a single cluster exceeds 50\% and the cluster has a mass less than 100 M$_\odot$, we reject the cluster. This restriction operates as a way to systematically clean up the tendency for \texttt{DBSCAN} to identify collections of field stars as a cluster despite the members being spatially and energetically separate. 

Once the clusters have been identified, we determine the half-mass radius $r_{\rm h}$ to give a sense of the compactness of each. We also record the mass of each cluster as well as what fraction of the total mass the cluster constitutes compared to that of all stars formed in the simulation at this point. 

The fragmenting of the cloud in the forced early massive star runs promotes the formation of distinct star clusters that move towards the edge and eventually out of the computational domain. Indeed, in the 100M run, all star clusters leave the computational domain by the end of the simulation. These distinct clusters in the forced runs seem to track the escaping gas and likely would not fall together in a collapse event, as we discuss further in Section \ref{sec:PrevWork}. 

\subsubsection{Cluster Properties}
Figure~\ref{fig:clust_grid_plot} shows that the fiducial run results in a relatively compact single cluster that represents over a third of all the initially available gas mass as well as 99\% of the stars that have formed in the simulation. This is in stark contrast to each of the early-forming massive star runs in which stellar material is broadly distributed across multiple spatially separated stellar associations,despite the systems being the same age and deriving from identical initial conditions. 
  
The fraction of mass contained within clusters also indicates increasing levels of disruption in cluster assembly in the early-forming massive star runs. In 50M, 97\% of all stellar mass is contained within the 4 identified clusters. In 70M, the mass fraction is slightly lower at 85\%. In 100M, only 46\% of stellar mass is contained within the identified clusters meaning over half of all stars formed in the run are either in very loosely bound associations, are lone field stars, or have left the computational domain. The 100M run has the most clusters rejected for insufficient mass or gravitational boundness with three, each of which can be picked out by eye in Figure \ref{fig:clust_grid_plot} as collections of grey points. The 70M and 50M runs both have a single cluster rejected. 
The clusters in the early-forming massive star runs are significantly less massive than the single cluster in the fiducial run, as expected since the fiducial run forms 2--3 times as much stellar material.

The half-mass radii included in Figure~\ref{fig:clust_grid_plot} show the relative compactness of the clusters. We calculate $r_{\rm h}$ from the center of mass of each cluster, excluding any gas that may be present within the cluster region. The single cluster in the fiducial run is compact, with half of its mass within 0.25 pc of the center of mass despite its outermost stars extending up to 2 pc away. Interestingly, the most massive clusters in each of the early-forming massive star runs (the blue cluster in each case) have the smallest values of $r_{\rm h}$ at 0.17, 0.21, and 0.18 pc for the 50M, 70M, and 100M runs respectively. 

We also calculate the boundness between the centers of masses of all clusters to the most massive cluster on the grid. In the 50M run, only the least massive identified cluster (red) is bound to the most massive cluster. In the 70M and 100M runs, the centers of masses for all identified clusters are unbound relative to the most massive cluster. Portions of the cluster in the fiducial run do break into two smaller subclusters of less than 100~M$_\odot$ at later times, but the centers of masses of those subclusters remain bound to the main fiducial cluster center of mass. The clusters formed in this work are more compact and lower in mass than the older clusters observed by \citet{kharchenko_global_2013}. However they are consistent with the six deeply embedded clusters observed by \citet{kuhn_spatial_2014}.
A summary of cluster comparison data is collected in Table \ref{tab:cluster_stats_2tff}.

\begin{table*}[!htb]
 \caption{Cluster statistics at $2\tau_{\rm ff}$.}
 \label{tab:cluster_stats_2tff}
 \begin{tabular}{lccccc} 
  \hline
  Run & \# clusters\tablenotemark{a} & Mass in Clusters\tablenotemark{b} & Frac Mass\tablenotemark{c} & $r_{\rm h}$ MMC\tablenotemark{d} & $E_{\rm bind}$ MMC\tablenotemark{e} \\
  &  & 10$^3$ $M_{\odot}$ & $M_{c}/M_{tot}$ & pc & $10^{46}$ erg\\
  \hline
  Fid & 1 & 3.6 & 0.99 & 0.25 & -140 \\
  50M & 4 & 1.4 & 0.97 & 0.17 & -12 \\
  70M & 4 & 0.86 & 0.85 & 0.21 & -4.2 \\
  100M & 2 & 0.62 & 0.46 & 0.18 & -3.8 \\
  \hline
 \end{tabular}
\tablenotetext{a}{Number of clusters on the grid identified by \texttt{DBSCAN} that pass mass and boundness criteria.} \tablenotetext{b}{The stellar mass contained within all identified clusters.} \tablenotetext{c}{The fraction of stellar material present in all present clusters relative to the total cumulative mass of stars that have formed.} \tablenotetext{d}{Half-mass radius of most massive cluster (MMC).}
\tablenotetext{e}{The binding energy of the most massive cluster (MMC).}
\end{table*}

The early-forming massive star in each run is not associated with any cluster.
This is not unusual since the star is able to rapidly drive away gas from its parent sink so that no subsequent star or cluster formation can occur in the massive star's vicinity. In addition, the rapid removal of gas from the center of the cloud allows the early-forming massive star and its parent sink to drift out of the cloud center since the gas gravitational potential is so severely disrupted. 
In the fiducial run at $2\tau_{\rm ff}$, ten very massive stars with M$>20$ M$_\odot$ are present in the identified cluster including 49, 61, and 67~M$_\odot$ stars. Of those, seven reside within $r_{\rm h}$, and six of those are within the 12.5\% mass-radius relation at just 0.084 pc. In the forced runs, the identified clusters have few or no stars with M$>20$ M$_\odot$ which is expected when considering the random sampling of the Kroupa distribution \citep{weidner_maximum_2006} of clusters with low total mass. 

We also analyze the most massive star in each cluster by finding the ratio of its mass to the approximate maximum stellar mass relation proposed by \citet{larson_stellar_2003}: $M_{\text{star}}\sim1.2M_{\text{cluster}}^{0.45}$. The fiducial run has a ratio of 1.41, the most massive star in the identified cluster is 1.41 times the \citet{larson_stellar_2003} relation. The four clusters in the 50M run have ratios of 0.65, 2.98, 1.56, and 1.45 in order of descending total cluster mass. The maximum mass relations of the 70M clusters run are 1.87, 1.83, 4.60, and 1.07. The 100M ratios are 1.06 and 1.17. In order to include the field stars not within identified clusters, we also take the ratio of the most massive star in the computational domain to the \citet{larson_stellar_2003} relation, using the sum of all stars as $M_{\text{cluster}}$. The ratios for the fiducial, 50M, 70M, and 100M runs are 1.28, 1.40, 2.16, and 0.74. This similarity between the fiducial and forced runs is expected as each sink is assigned its own mass list from which to form stars. A massive star that is forced to form early from one sink will not have a direct effect on the star masses formed by sinks elsewhere in the GMC. Overall, early-forming massive stars do not cause individual clusters or the entire star field to diverge from the \citet{larson_stellar_2003} relation.

In the fiducial run, the location and compactness of the distribution of massive stars is likely a result of early mass segregation \citep{binney_galactic_1987, mcmillan_dynamical_2007}. We find stars with $M >3M_\odot$ have $r_{\rm h} = 0.207$~pc while  for all stars $r_{\rm h} =0.247$~pc (excluding the $\sim67 M_\odot$ star several parsecs from the cluster center of mass yields $r_{\rm h} = 0.165$~pc for heavy stars). 

\section{Discussion} \label{sec:discussion}
\subsection{Previous Work}\label{sec:PrevWork}
The suppression of the global SFR and fragmentation of the initial GMC seen in our forced runs has been observed in other simulations, although under different circumstances. \citet{chen_effects_2021} found that variations in the initial gas density distribution of the GMC results in two modes of star cluster formation. GMCs with a steep power-law density distribution form a single massive cluster growing via direct gas accretion, similar to the cluster formed in our fiducial run. Nearly all of the mass in the fiducial cluster is due to star formation via its central sinks rather than through merger events with subclusters.
GMCs with top-hat gas distributions and otherwise identical properties fragmented into stellar subclusters similar to our forced massive runs. However, \citet{chen_effects_2021} observe these subclusters hierarchically assembling into central clusters, while we do not see significant evidence for hierarchical assembly in our forced massive star runs. 
The similar subcluster formation is likely due to the forced massive star forming very near to the density center of our GMC. The destructive effect of the stellar feedback occurring so early in the collapse of the cloud redistributed a significant portion of the GMCs initial mass to the outer sections of the cloud. The feedback acts as a source of internal pressure, preventing the gas from collapsing into the gravitational center and instead fragmenting into smaller, spatially distributed subclusters.

Similar to \citet{dale_before_2014}, we find that winds and ionizing feedback of deeply embedded massive stars disrupt the natal gas cloud. Our results also agree with the destruction of natal gas clouds found in the one-dimensional model \texttt{WARPFIELD}
\citep{rahner_warpfield_2019} and with the destructive effects reported in previous \texttt{Torch} simulations in which a very massive star forms \citep{wall_modeling_2020}.

In one of the simulations detailed in \citet{wall_modeling_2020}, a $\sim$97~M$_{\odot}$ star was born around 1.5 Myr after the onset of star formation in the main cluster. Its formation rapidly expelled gas from the cluster, halting local star formation, causing the least bound members of the cluster to be lost, but ultimately leaving the cluster intact. Similar results are exhibited in our fiducial run.

Our random stellar mass sampling technique mirrors the suggestion by \citet{weidner_maximum_2006} to produce a realistic IMF. A consequence of this technique is to trend to a stellar mass threshold which depends on the total mass of the star cluster \citep{larson_stellar_2003}. While our early-forming massive stars obviously diverge from the intended prescription, our desire is to probe the anomalous yet possible scenario that such a random sampling technique can produce very massive stars very early in the star formation history of a cluster. Still, our clusters adhere relatively well to the \citet{larson_stellar_2003} approximate maximum stellar mass relation. Comparing the most massive star in each cluster, seven of eleven are within a factor of 1.6 of the \citet{larson_stellar_2003} relation (see Sect.~\ref{subsec:expulsion}). Though there are clusters with divergent ratios it should be noted that these values are based on a snapshot of the clusters at $2\tau_{\rm ff}$ and so are subject to change as the clusters dynamically evolve and the massive stars shed mass as stellar winds.

\subsection{Limitations}
The initial conditions used in our runs are simplified to best test our hypothesis while also streamlining the simulation initialization as much as possible. Our turbulent spherical cloud neglects any of the complex dynamical interactions GMCs encounter in their pre-collapse phase including a galactic potential. Hence, we do not resolve any tidal effects the cloud may experience. Though, since the cloud evolves for only $2$~Myr prior to star formation, we assume the tidal effects of orbiting a galaxy to be negligible. We also start with a rather low virial parameter of 0.12 (with 0.5 being equilibrium). This low virial parameter effectively assumes an evolutionary history of global gravitational instability or galactic-scale collision at a scale larger than our simulations. 

Our model does not include pre-main-sequence feedback effects such as jets or accretion luminosity that would act during the main accretion phase. If we were to resolve the subgrid dynamics of stellar accretion, we would expect that the destructive effects of the forced massive stars should turn on even earlier to account for the star's pre-main sequence evolution, eventually reaching the ZAMS feedback that we see in our runs. The inclusion of instantaneous ZAMS feedback in our forced massive star runs results in a high likelihood that the forced massive star will remain unassociated with a cluster as its feedback immediately prevents its parent sink from forming any other stars. 
It is also possible that the pre-main-sequence feedback could further disrupt the hierarchical assembly process of the cluster and better halt the collapse of the initial cloud. That being said, massive star pre-main-sequence evolution is very fast, with the main accretion phase only lasting about $10^5$~yr \citep[e.g.][]{palla_star_1999}. Indeed our forced massive stars take only a few hundred thousand years to form from their parent sinks. Furthermore, the vast majority of the energy input from the forced massive stars is during their main-sequence and post-main-sequence evolution. 

The process through which massive stars form remains under debate \citep{tan_massive_2014,krumholz_formation_2015,motte_high-mass_2018}. Our models can not resolve the formation of individual massive stars. Instead we use a subgrid star formation model. This model identifies a region of gas that will continue to collapse, allows accretion into the region, and then forms stars chosen from the Kroupa IMF. Since we only model accretion into star forming regions at large scales of 0.5 pc, we remain agnostic to ongoing debates of the formation of individual stars such as the importance of fragmentation-induced starvation \citep{peters_limiting_2010} or monolithic collapse \citep{mckee_formation_2003}. Ultimately, our ability to form stars one by one and track their feedback, nuclear evolution, and dynamics allows us to probe the effects of massive stars on an actively forming star cluster. Also, the abnormally large mass of our early-forming massive stars may be unlikely to appear in reality, but within our computational model, the sink is able to accrete enough gas to justify such a star. Forcing a massive star to form early probes the extreme of what may be possible in a \texttt{Torch} simulation. Massive stars also preferentially form in multiples \citep{duchene_stellar_2013}, while this work does not include primordial binaries, \cite{cournoyer-cloutier_implementing_2021} successfully implemented a primordial binary model within \texttt{Torch} which will be included in future work. That said, \cite{cloutier_early_2023} finds that the early dynamical evolution of low mass clusters is not dominated by two- or few-body interactions but by the gravitational potential of the star forming region.

Higher resolution would have effects on how gas is accreted onto sinks, how stars are placed, and how stars interact with the gas. Higher maximum levels of refinement will require smaller sinks to form with smaller accretion and star placement regions, as well as capturing more detail of the massive star feedback interactions with gas. If one of our sinks is accreting material for a massive star on the list, no low mass star formation is allowed within its accretion radius. Higher grid refinement and smaller sinks could separate a region of dense gas into multiple star formation regions, allowing for other stars to form while one of the sinks remains in the accretion stage.  That said, once a very massive star is born, the effects on its surroundings are similar in low and high resolution runs. Feedback from massive stars will suppress star formation and expel gas even in simulations with higher maximum refinement levels.

Lastly, a significant fraction of the gas is ejected from the computational domain. This gas could recollapse, allowing for further star formation if it
leaves the computational domain with insufficient speed to escape the newly formed star cluster. Using results from the one-dimensional model \texttt{WARPFIELD} \citep{rahner_forming_2018}, we can check this possibility by examining the star formation efficiency and peak cloud density. Our runs have total star formation efficiencies of 0.13--0.39 and an initial peak cloud number density of 525 cm$^{-3}$.  These values exclude the possibility of a re-collapse event occurring.
We therefore assume that the gas that exits the computational domain will not recollapse into the star forming region and may be completely dispersed by later stellar feedback. The stars that form within our computational domain will continue to remain isolated from any gas that is ejected from the system, as well as any potential star formation that could occur in that gas.

\section{Conclusion} \label{sec:conclusion}
Using runs with identical initial conditions but with different times of formation of the first massive stars, we find that early-forming massive stars:
\begin{itemize}
    \item Globally unbind gas nearly 2~Myr earlier than our fiducial run.
    \item Reduce the global star formation efficiency by up to a factor of three and the average star formation efficiency per free-fall time by up to a factor of seven.
    \item Promote the formation of isolated stellar subclusters.
    \item Hinder the subclusters from collapsing into a single massive cluster.
\end{itemize}

The early-forming massive stars significantly disrupt the natal gas environment, globally unbinding gas nearly 2~Myr earlier than our fiducial run lacking the early formation of a massive star.
As a result, the massive stellar feedback evacuates the gas from the computational domain more rapidly in the runs with an early massive star, thus dramatically reducing the amount of star formation. While the gas in the fiducial run remains more centrally concentrated than in the runs with early-forming massive stars and the gas accretion rate onto sinks leading to the formation of stars is up to an order of magnitude higher, star formation in the fiducial run is still entirely quenched by around twice the free-fall time of the initial cloud, about 3~Myr after the onset of star formation. This termination of star formation occurs because of the global unbinding of the gas by feedback from later forming massive stars. 

In the early massive star runs, star formation still occurs for several million years after the gas reaches a globally unbound state. This is indicative of isolated subcluster formation, where small separated pockets of gas can continue to collapse and form stars even after most of the gas has been expelled. Among the three early massive star runs, the 100M and 70M runs expel more gas from the computational domain, produce and maintain less Jeans unstable gas, and ultimately form fewer stars than the 50M run. This trend is also seen in the comparison between the 50M and fiducial runs.

Before star formation is quenched in the fiducial run, around one-third of the initial gas is converted to stars, two to three times more than in the early massive star runs. We also find that early-forming massive stars cause star formation to occur in spatially separate and energetically unbound subclusters. In the fiducial run, in contrast, a single star cluster containing nearly all stars forms. At the same simulation time, the early massive star runs all have several clusters present but fewer stars associated with clusters, with the 100M run having less than half of all formed stars present in the two identified clusters. The most massive clusters in the early runs at $2\tau_{\rm ff}$ contain less than 40\% of the total stars formed and have masses of only a few hundred solar masses while the fiducial cluster has a mass of 3500 M$_\odot$. Although there likely will be a little more star formation occurring in the early runs as expelled gas continues to collapse outside of the computational domain, any last stars formed will likely not be associated with a larger single cluster and the gas is not likely to re-collapse to trigger a second round of star formation at the center of the cluster. 
Early-forming massive stars, in otherwise identical initial gas clouds, greatly disrupt the gas collapse, star formation, and cluster assembly processes.


\section*{Data Availability}
The data from the simulations and figures within this article will be shared on reasonable request to the corresponding author.

\begin{acknowledgments}
We acknowledge useful discussions of the stellar evolution code \texttt{SeBa} with S. Toonen.  
M.-M.M.L., S.L.W.M., and A.T. were partly supported by NSF grant AST18-15461. 
S.L.W.M. was also supported by NSF grant AST18-14772. 
A.T. was also supported through a NASA Cooperative Agreement awarded to the New York Space Grant Consortium. 
A.S. and C.C.-C. are supported by the Natural Sciences and Engineering Research Council of Canada. 
C.C.-C. also acknowledges funding from a Queen Elizabeth II Graduate Scholarship in Science and Technology (QEII-GSST). 
M.W. is supported by NOVA under project number 10.2.5.12. 
B.P. thanks the International Max Planck Research School for Astronomy and Cosmic Physics at the University of Heidelberg (IMPRS-HD) for their financial support. 
R.S.K. acknowledges support from the European Research Council via the ERC Synergy Grant ``ECOGAL'' (project ID 855130), from the Heidelberg Cluster of Excellence (EXC 2181 - 390900948) ``STRUCTURES'', funded by the German Excellence Strategy, from the German Research Foundation (DFG) in the Collaborative Research Center SFB 881 ``The Milky Way System'' (funding ID 138713538, subprojects A1, B1, B2, and B8), and from the German Ministry for Economic Affairs and Climate Action in project ``MAINN'' (funding ID 50OO2206). The Heidelberg group also acknowledges HPC resources and data storage supported by the Ministry of Science, Research and the Arts of the State of Baden-W\"{u}rttemberg (MWK) and DFG through grant INST 35/1314-1 FUGG and INST 35/1503-1 FUGG, and  computing time from the Leibniz Computing Center (LRZ) in project pr74nu. Simulations reported here were conducted on Cartesius; we acknowledge the Dutch National Supercomputing Center SURF grant 15520.  M.-M.M.L. thanks the Institut f\"ur Theoretische Astrophysik for hospitality during work on this paper.
\end{acknowledgments}

\facilities{Cartesius; SURF - Dutch National Supercomputing Center}

\software{\texttt{Torch} \citep{wall_collisional_2019,wall_modeling_2020}, \texttt{AMUSE} \citep{portegies_zwart_multiphysics_2009,portegies_zwart_multi-physics_2013,pelupessy_astrophysical_2013,portegies_zwart_astrophysical_2018}, \texttt{FLASH} \citep{fryxell_flash_2000}, \texttt{yt} \citep{turk_yt_2011}, numpy \citep{oliphant_python_2007}, scikit-learn \citep{pedregosa_f_scikit-learn_2011}, matplotlib \citep{hunter_matplotlib_2007}, HDF \citep{koranne_hierarchical_2011}.}

\bibliography{finalbib-no-doi.bib}{}
\bibliographystyle{aasjournal}


\end{document}